\documentclass[aps,pra,twocolumn,showpacs,amsmath,groupedaddress]{revtex4}  
\usepackage{graphicx}  
\usepackage{dcolumn}   
\usepackage{bm}        
\usepackage{amssymb}   
\usepackage{txfonts}

\begin{document}


\title{Quantum Zeno and anti-Zeno effect in atom-atom entanglement induced by non-Markovian environment}

\author{Yang Li}
\author{Hong Guo}\thanks{Author to whom correspondence should be
addressed. phone: +86-10-6275-7035, Fax: +86-10-6275-3208, E-mail:
hongguo@pku.edu.cn.} \affiliation{CREAM Group, State Key Laboratory
of Advanced  Optical Communication Systems and Networks (Peking
University)\\
and Institute of Quantum Electronics, School of Electronics
Engineering and Computer Science,
\\
and Center for Computational Science and Engineering (CCSE), Peking
University,
Beijing 100871, P. R. China}%

\date{\today}

\begin{abstract}
The dynamic behavior of the entanglement for two two-level atoms
coupled to a common lossy cavity is studied. We find that the speed
of disentanglement is a decreasing (increasing) function of the
damping rate of the cavity for on/near (far-off) resonant couplings.
The quantitative explanations for these phenomena are given, and
further, it is shown that they are related to the quantum Zeno and
anti-Zeno effect induced by the non-Markovian environment.
\end{abstract}

.

\maketitle

\section{Introduction}
Quantum entanglement plays an essential role in quantum information
science \cite{QCI}. However, entanglement is very fragile due to the
influence of the environment, which is recognized as the major
obstacle for the final application of quantum information
processing. For example, a phenomenon called ``entanglement sudden
death'' \cite{ESD1,ESD2,ESD3} shows that two initially entangled
qubits may disentangle completely in a finite time due to
spontaneous emission. So in order to keep the atomic entanglement
for a long time, spontaneous emission should be suppressed. Several
ways were proposed for this purpose. One way widely applied is to
place the qubits in a structured environment, say, microcavity
\cite{cavity2,cavity3} or in the photonic band gap of photonic
crystals \cite{PC1}, such that the qubits are separated from the
environment. Another way is to dynamically control the coupling
between the system and the environment, by, e.g., quantum Zeno
effect \cite{PM3}. In practice, people may combine the two ways
\cite{PM3,combine1,QZE2,IQZ1,IQZ2,IQZ3,IQZ4}. In fact, the two
methods have something in common.

In this paper, we investigate the entanglement dynamics of two
two-level atoms in a common cavity, a non-Markovian environment. We
find that when the atoms are on or near resonant with the cavity,
the speed of the disentanglement decreases as the quality factor of
the cavity decreases, and when the atoms are far-off-resonant with
the cavity, the speed of the disentanglement increases as the
quality factor of the cavity decreases. These phenomena can be
related to quantum Zeno \cite{EQZS1,EQZS2,QZE2} and anti-Zeno effect
\cite{IQZ1,IQZ2,IQZ3,IQZ4}.

\section{THEORETICAL Model}
Consider two spontaneously emitting two-level atoms with a common
zero-temperature bosonic reservoir, and in the Lamb-Dicke limit
\cite{DDH0}, the dipole-dipole interaction is negligible. Under the
rotating wave approximation (RWA), the Hamiltonian of this composite
system plus the reservoir is given by ($\hbar=1$)
\cite{PM3,DDH0,DDH} :
\begin{equation}
\label{Hamiltonian}
 H = H_0  + H_{{\mathop{ int}} } ,
\end{equation}
with
\[
H_0  = \omega _1 \sigma _ + ^{(1)} \sigma _ - ^{(1)}  + \omega _2
\sigma _ + ^{(2)} \sigma _ - ^{(2)}  + \int_{ - \infty }^\infty
{d\omega _k \omega _k b^\dag  (\omega _k )b(\omega _k )}  ,
\]
\[
H_{ int }  = (\alpha _1 \sigma _ + ^{(1)}  + \alpha _2 \sigma _ +
^{(2)} )\int_{ - \infty }^\infty  {d\omega _k g(\omega _k )b (\omega
_k )}  + {\rm h.c.}
\]
\noindent Here, $ \sigma _ \pm ^{(j)}$ and $\omega_{j}$ are the
inversion operators and transition frequency of the $j$th qubit ($j
= 1, 2$), and $b(\omega_k)$, $b^\dag(\omega_k)$ are the annihilation
and creation operators of the field mode of the reservoir. The mode
index $k$ contains several variables which are two orthogonal
polarization indices and the propagation vector $\vec k$. To measure
the coupling strength of the atoms to the cavity mode determined by
the atom's relative position in the cavity, we introduce the
dimensionless constant $\alpha_{j}$ \cite{PM3}. In the following
discussion, we introduce vacuum Rabi frequency $R = W(\alpha _1^2  +
\alpha _2^2 )^{1/2}$ and relative coupling strengthes $r_j  = \alpha
_j (\alpha _1^2  + \alpha _2^2 )^{ - 1/2}$ ($j=1,2$).

For an initial state of the form
\[ \left| {\psi (0)} \right\rangle
= (c_{10} \left| e \right\rangle _1 \left| g \right\rangle _2  +
c_{20} \left| g \right\rangle _1 \left| e \right\rangle _2 )\left| 0
\right\rangle _E ,
\]
since $[H,N] = 0$, where $ N = \int_{ - \infty }^\infty  {d\omega _k
 b^\dag  (\omega _k )b(\omega _k )}  + \sigma _ + ^{(1)}
\sigma _ - ^{(1)}  + \sigma _ + ^{(2)} \sigma _ - ^{(2)}  $, the
time evolution of the total system is confined to the subspace
spanned by the bases $ \{ \left| e \right\rangle _1 \left| g
\right\rangle _2 \left| 0 \right\rangle _E ,\left| g \right\rangle
_1 \left| e \right\rangle _2 \left| 0 \right\rangle _E ,\left| g
\right\rangle _1 \left| g \right\rangle _2 \left| {1_k  }
\right\rangle _E \} $:
\begin{eqnarray}
\label{eq2}
 \left| {\psi (t)} \right\rangle &=& c_1 (t)e^{ - i\omega _0 t} \left| e \right\rangle _1 \left| g \right\rangle _2 \left| 0 \right\rangle _E  + c_2 (t)e^{ - i\omega _0 t} \left| g \right\rangle _1 \left| e \right\rangle _2 \left| 0 \right\rangle _E
  \nonumber\\& &+
\int_{ - \infty }^\infty  {d\omega _k c_{\omega _k } (t)e^{ -
i\omega _k t} \left| g \right\rangle _1 \left| g \right\rangle _2
\left| {1_k  } \right\rangle } _E ,
\end{eqnarray}
where $\left| {1_k } \right\rangle _E $ is the state of the
reservoir with only one exciton in the $k$th mode. Here, we consider
the case in which the two atoms have the same Bohr frequency, i.e.,
$\omega_{1}=\omega_{2}=\omega_{0}$. Substituting Eq. (\ref{eq2})
into Schr$\rm \ddot{o}$dinger's equation and eliminating the
coefficients $c_{\omega _k } (t)$, one has

\addtocounter{equation}{1}
\begin{align}
\label{eq3a} \dot c_1 (t) =  - \int_0^t {dt_1 f(t - t_1 )\alpha _1
[\alpha _1 c_1 (t_1 ) + \alpha _2 c_2 (t_1 )]} , \tag{\theequation a}\\
\label{eq3b}\dot c_2 (t) =  - \int_0^t {dt_1 f(t - t_1 )\alpha _2
[\alpha _1 c_1 (t_1 ) + \alpha _2 c_2 (t_1 )]}  , \tag{\theequation
b}
\end{align}

\noindent where the correlation function takes the form:
\[f(t - t_1 ) = \int_{ - \infty
}^\infty  {d\omega _k J(\omega _k )} e^{ - i(\omega _k  - \omega _0
)(t - t_1 )}
.
\]
As discussed in \cite{PM3}, there is a subradiant state, $\left|
\psi_{-} \right\rangle = r_2 \left| e \right\rangle_1 \left| g
\right\rangle_2 - r_1 \left| g \right\rangle_1 \left| e
\right\rangle_2 $, which does not decay in time, and the only
relevant time evolution is the superradiant state $\left| \psi_{+}
\right\rangle = r_1 \left| e \right\rangle_1 \left| g
\right\rangle_2 + r_2 \left| g \right\rangle_1 \left| e
\right\rangle_2$.

For the case of two atoms interacting with a cavity field in
presence of cavity losses, the spectral density function takes the
form
\begin{equation}
\label{spectrum} J(\omega _k ) = {W}^2 \lambda /\pi [(\omega _k  -
\omega _c )^2 + \lambda ^2 ],
\end{equation}
where $W$ is the transition strength, $\omega_c$ is the center of
the spectrum, and $2\lambda$ is the full width at half maximum
(FWHM) of the spectral function. By employing Fourier transform and
residue theorem, we get the explicit form $f(t - t_1 ) = {W}^2 e^{ -
\lambda \left| {t - t_1 } \right|}$, where the quantity $1/\lambda$
is the reservoir correlation time.

Using Laplace transform, we get the solutions of Eqs. (\ref{eq3a})
and (\ref{eq3b}):

\addtocounter{equation}{1}
\begin{align}
\label{eq5a} c_1 (t) = r_2 \beta_{-} + r_1 \beta_{+} \varepsilon(t) , \tag{\theequation a}\\
\label{eq5b} c_2 (t) = - r_1 \beta_{-} + r_2 \beta_{+}
\varepsilon(t) , \tag{\theequation b}
\end{align}

\noindent where $\beta_{\pm} = \left\langle {{\psi _ \pm  }}
 \mathrel{\left | {\vphantom {{\psi _ \pm  } {\psi (0)}}}
 \right. \kern-\nulldelimiterspace}
 {{\psi (t)}} \right\rangle$, $\varepsilon(t) = (s_{+} + \lambda +i \delta) e^{s_{+}
t}/(s_{+} - s_{-})-(s_{-} + \lambda + i \delta) e^{s_{-} t}/(s_{+} -
s_{-})$, and $s_{\pm}$ are the roots of the equation for $s$: $s^2 +
(\lambda + i \delta) s + R^2=0 $, where $\delta = \omega_c -
\omega_0$ is the detuning. In the $ \{ \left| e \right\rangle _1
\left| e \right\rangle _2 ,\left| e \right\rangle _1 \left| g
\right\rangle _2 ,\left| g \right\rangle _1 \left| e \right\rangle
_2 ,\left| g \right\rangle _1 \left| g \right\rangle _2 \} $ basis,
the reduced density matrix of the two atoms is given by:
\begin{equation}\label{rhoa}
\rho _a (t) = \left( {\begin{array}{*{20}c}
   0 & 0 & 0 & 0  \\
   0 & {\left| {c_1 (t)} \right|^2 } & {c_1 (t)c_2^* (t)} & 0  \\
   0 & {c_2 (t)c_1^* (t)} & {\left| {c_2 (t)} \right|^2 } & 0  \\
   0 & 0 & 0 & {1 - \left| {c_1 (t)} \right|^2  - \left| {c_2 (t)} \right|^2 }  \\
\end{array}} \right)
.\end{equation}

\noindent The entanglement of the two atoms can be evaluated by
concurrence $C(t)$ \cite{Concurrence}. For $\rho_a$ [Eq.
(\ref{rhoa})], its concurrence can be derived from
\cite{Concurrence}, as
\begin{equation}
\label{concurrence} C(t) = 2\left| {c_1 (t)c_2^* (t)} \right| =
2\left| {c_1 (t)} \right|\left| {c_2 (t)} \right| .
\end{equation}

\section{Numerical Results}
We focus on the concurrence as a function of time $t$ in
weak-coupling regime, $W < \lambda /2$. In this regime, the
concurrence of the two atoms undergoes nearly irreversible
exponential decay. Similar behaviors mentioned below take place for
strong-coupling regime in the time scale of Rabi oscillation.

We compare the entanglement dynamics of the two atoms initially in
the maximal entanglement states  $\left| {\varphi _ +  }
\right\rangle = \frac{1}{{\sqrt 2 }}(\left| e \right\rangle _1
\left| g \right\rangle _2  + \left| g \right\rangle _1 \left| e
\right\rangle _2 ) $ for three different values of full width at
half maximum (FWHM) of the spectral function, namely, $2\lambda =
10, 16, 20$.

As mentioned in \cite{Position}, it is hard to exactly control the
position of the atom in the optical cavity. But through numerical
stimulation, we find that different values of $r_1$ show
qualitatively similar behaviors. So, for simplicity, we focus on the
case of equal coupling parameters, i.e., $r_1 = r_2 = \sqrt{2}/2$.

\begin{figure}
\centering
\includegraphics[height=0.40\textheight,width=0.4\textwidth]{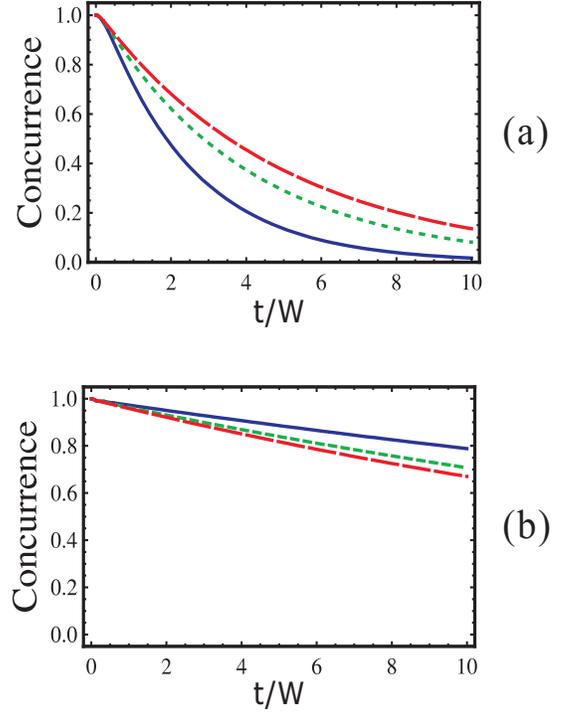}
\caption{\small (Color online) Time evolution of the concurrence,
with (a) $\delta=0$ and (b) $\delta=20$, both with the initial state
being the maximal entanglement state $\left| {\varphi_{+} }
\right\rangle $, $W=1$, and $r_1=r_2$, for the cases of (i)
$\lambda=5$ (solid blue curve), (ii) $\lambda=8$ (dotted green
curve), (iii) $\lambda=10$ (dashed red curve) .} \label{fig1}
\end{figure}

As in FIG. \ref{fig1}(a), the atoms are on resonance with the center
of the spectrum, $\delta = 0$. The concurrence decreases
monotonically down to zero at the beginning. An interesting
phenomenon is that the speed of disentanglement decreases as
$\lambda$ increases. Similar behavior happens when the atoms are
near resonance with the center of the spectrum, $\delta \ll
\lambda$. In fact, this phenomenon is related to the environment
induced quantum Zeno effect \cite{EQZS1,EQZS2,QZE2}. However, When
the atoms are far off-resonant with the center of the spectrum,
$\delta \gg \lambda$, for example, as shown in FIG. \ref{fig1}(b),
where we choose $\delta = 20$, the speed of disentanglement
increases as $\lambda$ increases. This latter phenomenon is related
to the anti-Zeno effect \cite{IQZ1,IQZ2,IQZ3,IQZ4}.

In fact, as mentioned in \cite{Li}, if the coupling strengths of the
two atoms to the field are different and the dipole-dipole
interaction is not negligible, there are no asymptotic entanglement,
which means that even $\left| {\psi _ -  } \right\rangle $ will
disentangle completely. In these cases, our numerical stimulation
shows that similar phenomena mentioned above happen.

\begin{figure}
\centering
\includegraphics[height=0.20\textheight,width=0.4\textwidth]{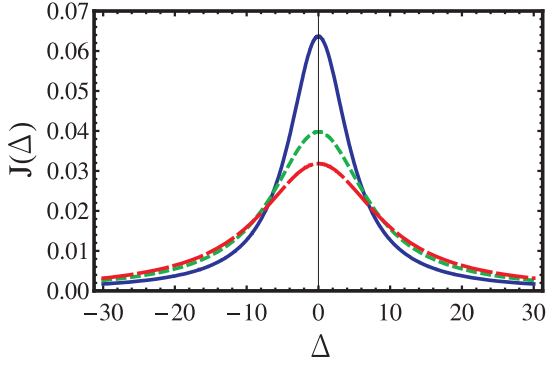}
\caption{\small (Color online) The spectrum of the field in the
cavity, $J(\Delta) = W^2 \lambda/\pi(\Delta^2 + \lambda^2)$, $W=1$,
for the cases of (i) $\lambda=5$ (solid blue curve), (ii)
$\lambda=8$ (dotted green curve), (iii) $\lambda=10$ (dashed red
curve) .} \label{fig2}
\end{figure}

We can give an intuitive explanation for these phenomena. As we can
see in FIG. \ref{fig2}, the center part of the spectrum decreases
monotonically as $\lambda$ increases, while the parts which are far
from the center increases as $\lambda$ increases. We can prove that
the short-time behavior of the disentanglement is determined by the
modes of the spectrum which are on resonance with the atoms: the
speed of the disentanglement decreases (increases) as the density of
these modes decreases (increases).

Similar to \cite{QZE2,IQZ3}, we define $P(t) \equiv \left|
{\varepsilon (t)} \right|^2 \equiv e^{ - Rt}$, where $R$ is the
effective decay rate. From Eqs. (\ref{eq5a}), (\ref{eq5b}) and
(\ref{concurrence}), we can see that $P(t)$ and $R$ can describe the
concurrence to some extent. For short-time behavior, in the
first-order approximation, one yields \cite{IQZ3}:
\addtocounter{equation}{1}
\begin{align}
R = 2\pi \int_{-\infty}^\infty  {d\omega J(\omega )F(\omega )}, \tag{\theequation a}\\
F(\omega ) = \frac{t}{2 \pi }  {\rm sinc}^2 \left(\frac{{\omega  -
\omega _0 }}{2}\right)t. \tag{\theequation b}
\end{align}
Since $F(\omega)$ is a sharply steep function of $\omega$ around
$\omega_0$, if the width of $F(\omega)$ is much smaller than that of
$J(\omega)$, $R$ is mainly determined by $J(\omega_0)$. In this
paper, the form of $J(\omega)$ is given by Eq. (\ref{spectrum}).
Then a detailed analysis shows that if $\lambda$ is larger than
$\omega_0 - \omega_c$, $J(\omega_0)$ is a monotonic decreasing
function with respect to $\lambda$, otherwise, it is a monotonic
increasing function with respect to $\lambda$. Since $R \propto
J(\omega_0)$, these results also go to $R$.

\section{Environment induced quantum Zeno effect and anti-Zeno effect}
In order to relate these phenomena to environment induced quantum
Zeno and anti-Zeno effect, we make a transform on the original
Hamiltonian [Eq. (\ref{Hamiltonian})] as in \cite{PM2,Li}:
\begin{equation}
a = (\lambda /\pi )^{1/2} \int_{ - \infty }^\infty  {d\omega _k
\frac{{b(\omega _k )}}{{\omega _k  - \omega _c  - i\lambda }}} ,
\end{equation}
\begin{eqnarray}
c(\Delta ) &=& (\lambda /\pi ) {\rm P.V.} \left\{\int_{ - \infty
}^\infty {d\omega _k \frac{{b(\omega _k )}}{{(\omega _k  - \Delta
)(\omega _k
- \omega _c  - i\lambda )}}}\right\} \nonumber\\
& &+ \int_{ - \infty }^\infty {d\omega _k \frac{{\omega _k  - \omega
_c }}{{\omega _k  - \omega _c - i\lambda }}b(\omega_k )} ,
\end{eqnarray}
we get
\begin{eqnarray}
H = H_{0}' + H_{int}' +H_{damping},
\end{eqnarray}
\begin{eqnarray}
H_0 '  = \omega _1 \sigma _ + ^{(1)} \sigma _ - ^{(1)}  + \omega _2
\sigma _ + ^{(2)} \sigma _ - ^{(2)}  + \omega _c a^\dag  a
 ,
\end{eqnarray}
\begin{equation}
H_{ int }'  = W[(\alpha _1 \sigma _ + ^{(1)}  + \alpha _2 \sigma _ +
^{(2)} )a + (\alpha _1 \sigma _ - ^{(1)}  + \alpha _2 \sigma _ -
^{(2)} )a^\dag],
\end{equation}
\begin{eqnarray}
H_{ damping } &=& \int_{ - \infty }^\infty  {d\Delta \Delta c^\dag
(\Delta )c(\Delta )} \nonumber\\
& &+ (\lambda /\pi )^{1/2} \int_{ - \infty }^\infty {d\Delta [a^\dag
c(\Delta ) + ac^\dag  (\Delta )]},
\end{eqnarray}
\noindent where $a^\dag$ and $a$ are the creation and annihilation
operators for the discrete quasimode,  with the frequency
$\omega_c$, and $c^\dag (\Delta )$, $ c(\Delta)$ are those for the
continuum quasimodes with the frequency $\Delta$.

It can be seen, from the above Hamiltonian, that the composite
two-atom system only interacts with one discrete mode and their
coupling coefficient is just the transition strength $W$. The
discrete mode interacts with the continuum modes and their coupling
strength only contains the width ($\lambda$) of the Lorentzian
spectral density which is a constant. This  means that if we let the
two atoms and the discrete mode be a new system and the continuum
modes be the reservoir, the behavior of the new system is exactly
Markovian. As discussed in \cite{PM2}, the discrete mode is
appropriate for describing the electromagnetic field inside the
cavity, and the continuum modes are appropriate for describing the
field outside the cavity.

In terms of quantum measurement theory \cite{Measure1,Measure2,QO1},
the continuum quasimodes $c(\Delta)$ represent the environment, the
discrete quasimode $a$ represent the meter, and the two atoms are
the system. By interpreting the environment as a photoelectron
counter, the coupling strength of interaction between the discrete
quasimode and the continuum quasimodes relates to the rate of photon
counting $\lambda$. That means the environment acts as an observer
of the pointer states of discrete quasimode. Although, in our model,
$\left| 1 \right\rangle_a$ and $\left| 0 \right\rangle_a$ are not
perfect pointer states for discrete quasimode, the observation by
the environment destroys the coherence between them. We can
interpret $\tau = \lambda ^{-1}$ as the effective measurement
interval by the environment. Assume that the two atoms and the
discrete quasimode are in the state $\left| {\varphi_{+} }
\right\rangle_{12} \left| {0 } \right\rangle_a $ in the beginning.
For simplicity, we choose $\alpha_1=\alpha_2$, then because of the
interaction between the atoms and the quasimode, at a subsequent
time $\tau$, they evolve to the state $\left| {\varphi_{+} }
\right\rangle_{12} \left| {0} \right\rangle_a + W \tau \left| {gg}
\right\rangle_{12} \left| {1} \right\rangle_a$, and the probability
of the transition $\left| {\varphi_{+} } \right\rangle_{12} \left|
{0} \right\rangle_a \to \left| {gg} \right\rangle_{12} \left| {1}
\right\rangle_a$ is $(W \tau)^2$. In the meantime, the coherence of
this superposition is destroyed by the environment. Then the
probability of the transition after a time $\Delta t \gg \tau$ is $P
\approx (W \tau)^2 \Delta t/\tau = W^2 \tau \Delta t$. So, we can
see that this probability is proportion to $\lambda^{-1}$. It
explains the phenomenon that the speed of disentanglement decreases
as the strength of damping increases. In terms of quantum
measurement theory, this means that the observation made by the
environment suppress the disentanglement, which is just quantum Zeno
effect. When $\omega_0 - \omega_c$ is much larger than $\lambda$,
that means the frequency of the observation is very small, and in
the language in \cite{IQZ2,IQZ4}, we can say that the observations
are made in the anti-Zeno regime, so the observations accelerate the
disentanglement. A similar discussion about the Zeno and anti-Zeno
effect for nonresonant systems is given in \cite{IQZ4}, where the
observations are not made by the environment.

In conclusion, we extend the study of the entanglement dynamics of
two atoms in a common cavity. We find that the speed of the
disentanglement of the two atoms is a decreasing (increasing)
function of the damping rate of the cavity when the atoms are
on/near resonance (far off resonance)  with the center of the cavity
modes. We give a quantitative explanation for these phenomena, and
relate them to quantum Zeno and anti-Zeno effect induced by the
environment. These results are helpful for understanding the related
experimental phenomena and for the practical engineering of
entanglement in the future.

The authors appreciate Ting Yu for many fruitful discussions. This
work is supported by the Key Project of the National Natural Science
Foundation of China (Grant No. 60837004).

\end{document}